\documentclass{sig-alternate}
\usepackage{url}
\usepackage{multirow}

\newcommand{\argmax}{\operatornamewithlimits{argmax}}

\usepackage{times}
\usepackage{latexsym}
\usepackage{amsmath}
\usepackage{multirow}
\usepackage{graphicx}
\usepackage{multirow}
\usepackage{color}
\usepackage{url}
\usepackage{enumitem}
  {\begin{itemize}[topsep=0pt, partopsep=0pt] \footnotesize%
    \setlength{\itemsep}{0pt}%
    \setlength{\parskip}{0pt}%
    }%
  {\end{itemize}}
\usepackage{subfigure}

  {\begin{itemize}[topsep=10pt, partopsep=0pt] \footnotesize%
    \setlength{\itemsep}{2pt}%
    \setlength{\parskip}{0pt}%
    }%
  {\end{itemize}}

\begin{document}

\title{Early Stage Influenza Detection from Twitter}
\numberofauthors{2} 
\author{
\alignauthor
Jiwei Li\titlenote{Dr.~Trovato insisted his name be first.}\\
       \affaddr{School of Computer Science}\\
      \affaddr{Carnegie Mellon University}\\
       \affaddr{Pittsburgh, PA 15213}\\
       \email{bdlijiwei@gmail.com}
\alignauthor
Claire Cardie\titlenote{The secretary disavows
any knowledge of this author's actions.}\\
       \affaddr{Department of Computer Science}\\
       \affaddr{Cornell University}\\
       \affaddr{Ithaca, NY 14850}\\
       \email{cardie@cs.cornell.edu}
}

\maketitle
\begin{abstract}
Influenza is an acute respiratory illness that occurs virtually every year and results in substantial disease, death and expense. 
Detection of Influenza in its earliest stage would facilitate timely action that could
reduce the spread of the illness. 
Existing systems such as CDC and EISS which try to collect diagnosis data, are almost entirely manual, resulting in
about two-week delays for clinical data acquisition.
Twitter, a popular microblogging service, provides us with a perfect source
for early-stage flu detection due to its real-time nature.
For example, when a flu breaks out, people that get the flu may post related tweets which enables the detection of the flu breakout promptly.
In this paper, we investigate the real-time flu detection problem on Twitter data by proposing Flu Markov Network (Flu-MN): a spatio-temporal unsupervised Bayesian algorithm based on a 4 phase Markov Network, trying to identify the flu breakout at the earliest stage.
We test our model on real Twitter datasets from the United States along with baselines in multiple applications, such as real-time flu breakout detection, future epidemic phase prediction, or Influenza-like illness (ILI) physician visits.
Experimental results show the robustness and effectiveness of our approach. We build up a real time flu reporting system based on the proposed approach, and we are hopeful that it would help government or health organizations in identifying flu
outbreaks and facilitating timely actions to decrease unnecessary mortality.

\end{abstract}




\section{Introduction}
\label{intro-sec}
Influenza is an acute respiratory illness that occurs every year. Approximately 10-15$\%$ of people get influenza each year and such disease results in up to 50 million illnesses and 500,000 deaths in the world every year. Detection of Influenza in its earliest stage would facilitate timely action that could
reduce the spread of the illness and decrease the number of deaths. 
Indeed, early stage influenza detection has gained increasing importance recently due to the threat of newly emerging infections such as SARS, H5N1 and H5N9.
Influenza surveillance systems have been established via the European Influenza Surveillance
Scheme (EISS) in Europe and the Center for Disease Control (CDC) in the US to collect data
from clinical diagnoses. 
Unfortunately, these systems are almost entirely manual, resulting in
about two-week delays for clinical data acquisition.
Public health authorities need to forecast the breakout of Influenza at the earliest time to ensure effective preventive
intervention, giving rise to the increasing need for efficient sources of data
for forecasting.

\begin{figure}
\centering
\includegraphics[ width=3in,natwidth=610,natheight=642]{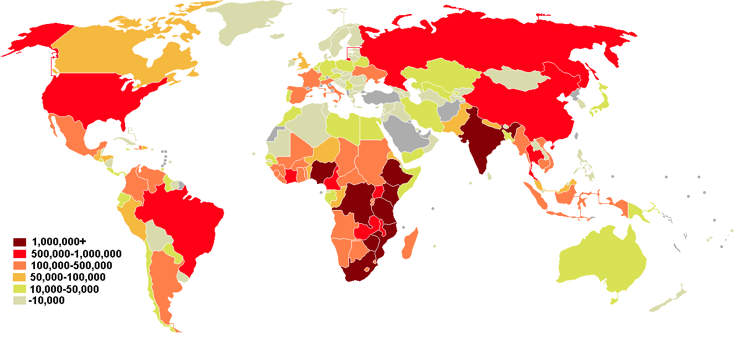}
\caption{World Flu map.}\label{fig:CDC}
\end{figure}
\begin{figure}
\centering
\includegraphics[ width=3in,natwidth=610,natheight=642]{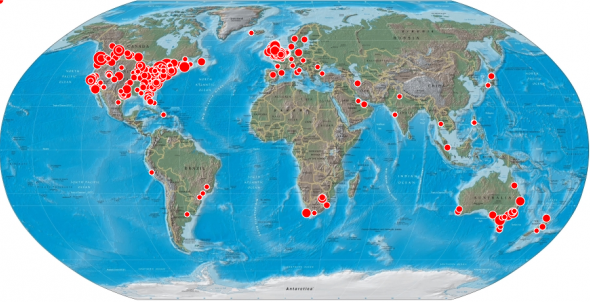}
\caption{World Twitter user map.}\label{fig:CDC}
\end{figure}
In an attempt to create new sources of flu reporting systems, early researchers used different types of influenza signals for surveillance, such as telephone calls \cite{espino2003telephone} or drug sales \cite{magruder2003evaluation}. The extensive development of world wide web gives rise to novel sources for flu detection.
Common approaches on web-based influenza
surveillance usually make use of click-through
data from search engines \cite{eysenbach2006infodemiology} such as
counting search queries submitted to a medical website \cite{hulth2009web}, 
visitors to health websites \cite{johnson2004analysis} or
clicks on a search keyword advertisement \cite{polgreen2008using}. Another important click-based flu reporting system is the
famous Google's flu trends service \cite{cook2011assessing,ginsberg2008detecting}\footnote{\url{http://www.google.org/flutrends/}}. 
Ginsberg et al. \cite{ginsberg2008detecting} introduced the linear model in Google's flu trend that links the 
influenza-like illness (ILI) visits with ILI-related search query to detect influenza epidemics.

Twitter, a popular microblogging service, potentially can provide a good source
for early stage flu detection due to its real-time nature.
When flu breaks out, infected Twitter users might post related tweets in a timely way, e.g. ``stuck at home with the flu".  These, in turn, can be regarded as indicators of Influenza and enable the detection of flu
occurrence promptly.
Due to the real time nature of Twitter, it has been used for many real-time applications such as earthquake detection \cite{sakaki2010earthquake}, 
public health tracking \cite{collier2012uncovering,paul2011you} and has already been proved to be quite useful for flu detection \cite{achrekar2011predicting,culotta2010towards}.
Lamb et al. \cite{lamb2013separating} explored multiple features in a supervised learning framework and try to select out tweets indicating flu around. Achrekar et al. \cite{achrekar2011predicting} created a system that can monitor flu-related tweets by extracting relevant location and user demographic information.
Other approaches for flu detection on Twitter involve supervised classification \cite{aramaki2011twitter,culotta2010towards}, 
unsupervised models, \cite{paul2011you}, keyword counting\footnote{The DHHS competition relied solely on keyword counting.
\url{http://www.nowtrendingchallenge.com/}}, or tracking geographic illness trend \cite{sadilek2012modeling}.

\begin{figure}
\centering
\includegraphics[width=2.4in]{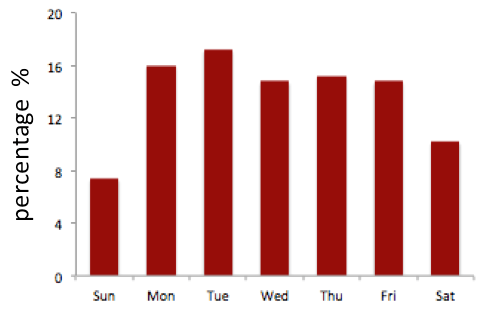}
\caption{Average flu related tweets rate within the week based on twitter data from Jun. 08 to Jun. 09.}\label{mn}
\end{figure}

Existing approaches (including Google's flu trend) usually use the $qualitative$ "sharp increase" in the amount of flu related signals (e.g clicks or tweets) as the flu breakout signals, but seldom explore the algorithm to quantitatively predict the exact time in real-time  when government or health organizations should send out alarm for flu breakout. To be specific, in twitter, most existing works \cite{aramaki2011twitter,culotta2010towards,lamb2013separating,paul2011you} pay more attention to how well tweet data can fit real world CDC data, but seldom explored the algorithm for flu breakout detection. In addition, existing works on twitter flu prediction suffer the following two limitations: (1) {\it spatial information} is seldom considered. Spatial dependency is important in early-stage flu detection, especially in the case where current location does not show clear clue for flu breakout but most of its neighbors do. This has been demonstrated in many works in Epidemiology \cite{aschwanden2004spatial,chowell2007estimation,duryea1999population,mollison1995epidemic}. (2) {\it daily effect} corresponds to fluctuation of the number of tweets posted based on what day it is within a week, or whether it is holiday etc. Such effect is seldom considered in existing works. As we can see in Figure \ref{mn}, users publish tweets about flu almost twice as much on Monday as they do on Sunday. Neglecting such effects would result in mistakes in flu reporting.

To address these problems, in this paper, we introduce an unsupervised approach called Flu Markov Network (Flu-MN) for early stage flu detection by framing spatial and temporal information in a unified Markov Network.  
Our approach assumes twitter users as "sensors" and collective tweets containing flu keywords as early indicators and robust predictors of influenza.
Spatial and temporal information are modeled in a four-phase Markov switching model, i.e. {\bf non-epidemic phase (NE)}, {\bf rising epidemic phase (RE)}, {\bf stationary epidemic phase (SE)} and {\bf declining epidemic phase (DE)}, where the dynamics of {\bf NE} and {\bf SE} are characterized by small changes
in the number of flu-indicator tweets, modeled as a {\it Gaussian process} based on {\it daily effect}, 
while the dynamics of {\bf RE} and {\bf SE} are modeled as an {\it autoregressive process} to account 
for the {\it auto dependent} changes
expected in the number of flu-indicator tweets.
Unlike existing approaches for spatio-temporal data analysis such as linear regressive process \cite{cressie2011statistics,waller1997hierarchical} or Markov Random Field \cite{lavine1999markov}, which require labeling data for weight training (which would be hard for our case), our approach is totally unsupervised.
We further used a 
Markov Chain Monte Carlo (MCMC) methods for parameter inference. 
We conduct experiments in multiple real-time applications (e.g. flu breakout prediction, ILI physician visits prediction). Our research
results verify (1) the effectiveness of using real-time twitter data for the task of flu detection (2) the good performances of our proposed model.
In sum, the main contributions of this paper are as follows:
\begin{enumerate}
\item We introduce an unsupervised Bayesian framework based on a spatio-temporal Markov Network algorithm for early stage flu detection on twitter. Our algorithm can capture flu outbreaks more promptly and accurately compared with baselines.
\item Based on our proposed algorithm, we create a real-time flu surveillance system.
From the system, we verify that twitter serves a better flu prediction resource than other types of signals, say query-click on search engine (See Section 4.4). 
\end{enumerate}

The rest of this paper is organized as follows: In the next section, we present our approach for the identification of flu-indicator tweets.
Our Flu-MN model is illustrated in Section~\ref{flu-hmm-sec}. 
We describe the creation of our dataset in Section \ref{experiments-1}.
Experimental results are illustrated in Section~\ref{experiments-2}.
Section \ref{rel-work-sec} is devoted to an explanation of related work. We conclude this paper in Section~\ref{conclusion-sec}.
\begin{figure}
\includegraphics[width=3in]{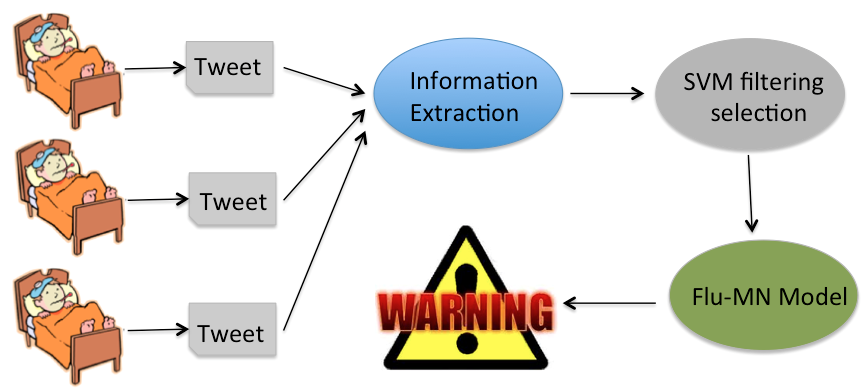}
\caption{Pipeline for Flu detection on Twitter.}
\end{figure}

\section{Model}
\label{flu-hmm-sec}
\begin{figure} 
\includegraphics[width=3in]{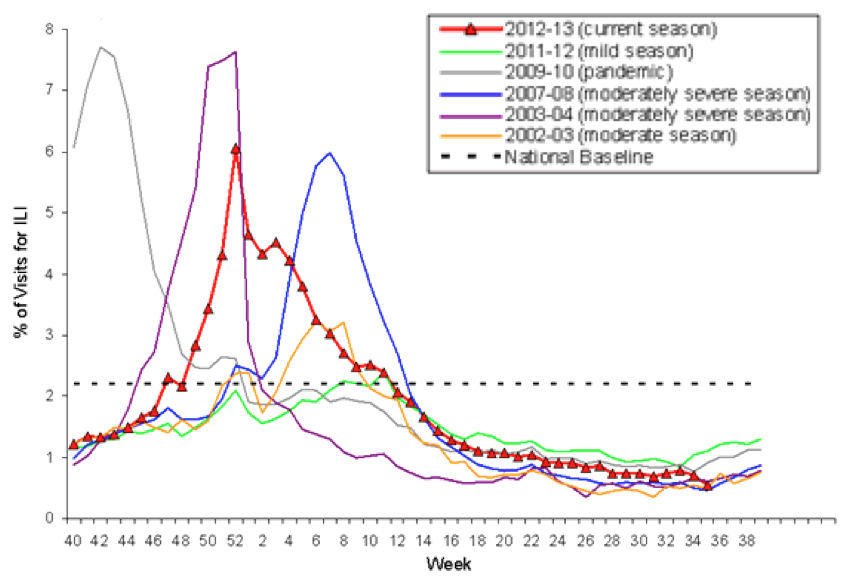}
\caption{Percentage of Visits for Influenza-like illness (ILI) reported by the US. Outpatient Influenza-like illness Surveillance Network.}\label{fig5}
\end{figure}
For early stage flu detection, we use a probabilistic graphical Bayesian approach based on Markov Network. We first describe the details of our model and then the inference procedure.

Suppose we collect flu-related tweet data from $N$ locations. In this paper, each location denotes a State in the US. For each location $i\in [1,N]$, we segment the data into a time series. $Z_{i,t}$ denotes the phase location $i$ takes on at time $t$. $Z_{it}=0,1,2,3$ correspond to the phases {\bf NE}, {\bf RE}, {\bf SE} and {\bf DE} respectively. $Y_{i,t}$ is the observant variable, which denotes the number of flu related tweets at time $t$, location $i$.
$\Delta Y_{i,t}=(Y_{i,t}-Y_{i,t-1})/Y_{i,t-1}$.
For location $i$, 
$Neigh(i)$ denotes the subset containing its neighbors.
We simplify the model by only considering bordering States in $Neigh(i)$.

\subsection{Daily Effect}
The first thing we should consider is what we call {\it daily effect}, which describes the people's inclination to post flu related tweets on each day within the week. As we observe from Figure \ref{mn}\footnote{\url{http://www.cdc.gov/flu/weekly/}}, more users publish flu-related tweets on weekdays than on weekends. 
To capture the dynamics of {\t daily effect},
we use a $Gaussian$ distribution.  Let $k=day(t)$ denote the day within the week (e.g. Monday, Tuesday). For time $t$, the observations $\Delta Y_{i,t}$ are modeled as a drawn from a $Gaussian$ distribution with mean $L_k$ and variance $\delta_k^2$:
\begin{equation}
\Delta Y_{i,t}\sim N(L_k, \delta_k^2)
\end{equation}
Based on Bayesian perspective, we give a $Gaussian$ prior for $L_k$ and an {\it Inverse-$\chi^2$} prior for $\delta_k^2$ based on previous data. $W_k$ denotes the mean of $\Delta Y_{i,t}$ with $day(t)=k$ within that time period and $G_k^2$ denotes the variance based on previous data from Jun.08 to Jun.09. 
\begin{equation}
L_k\sim N(W_k,\phi_k^2)
\end{equation}
\begin{equation}
 \delta_k^2\sim Inv-\chi^2(\mu_k,G_k^2)
\end{equation}

\subsection{Flu-MN}
The key of the flu detection task is to detect the transition time from non-epidemic phase to epidemic phase. 
As we examine the actual data at Figure \ref{fig5}, we can clearly identify non-epidemic phase ({\bf NE}), rising-epidemic phase ({\bf RE}) and declining-epidemic phase ({\bf DE}) in flu dynamics in each year. For some years (i.e. 2009), we can also find a stationary-epidemic phase ({\bf SE}) where the number of ILI physician visits approximately keeps at a fixed high value.
Hence, we segment each time into four phases {\bf NE}, {\bf RE}, {\bf SE} and {\bf DE}\footnote{Such 4 phases are called lag, log, stationary and death phase in bacteriology.}.
Such description of four-phase population characteristics can be found in many areas such as economy \cite{cochrane2005time}, bacteriology\footnote{\url{http://en.wikipedia.org/wiki/Bacterial_growth}}, population ecology \cite{odum1971fundamentals} or anthropology \cite{bateson2000steps}.
\paragraph{Spatial and Temporal dependency}
We model the spatial and temporal information in a unified Markov Network, where the phase for location $i$ at each time is not only dependent upon its previous phase, but its neighbors.
In this work, for simplification, we only treat bordering States as neighbors.
Since the influence from non-bordering States can be transmitted through bordering ones, such simplification makes sense and experimental results also demonstrate this point. We use a {\it Generalized Linear Model} to integrate the temporal and spatial information in a unified framework. For location $i$ at time $t$, the probability that $Z_{i,t}$ takes on value $Z$ is illustrated as follows: 
\begin{equation}
\begin{aligned}
&Pr(Z_{i,t}|Z_{i,t-1},Z_{j,t,+1}~j\in Neigh(i))=\\
&\frac{\exp(\Theta_{Z_{i,t-1},Z_{i,t}}+\Theta_{Z_{i,t},Z_{i,t+1}}+\sum_{j\in Neigh(i)}\Psi_{Z_{j,t},Z_{i,t}})}{\sum_{z}\exp(\Theta_{Z_{i,t-1},Z_{i,t}}+\Theta_{Z_{i,t},Z_{i,t+1}}+\sum_{j\in Neigh(i)}\Psi_{Z_{j,t},Z_{i,t}})}
\end{aligned}
\label{eq1}
\end{equation}
where $\Theta$ and $\Psi$ respectively correspond to parameters that control temporal and spatial influence. Take $\Theta$ for example, it is a $4\times 4$ metrics. $\Theta_{i,j}$ represents how likely phase $Z_{i,t}$ is to take value $j$ at time $t$ given that phase $Z_{i,t-1}$ takes on value of $i$ at $t-1$. We give a non-informative $Gaussian$ prior for each element in $\Theta$ and $\Psi$:
\begin{equation}
\Theta_{i,j}\sim N(0, \sigma^2_{i,j})~~~~~\Psi_{i,j}\sim N(0,\Phi^2_{i,j})
\end{equation}

\begin{figure}
\centering
\includegraphics[width=3in]{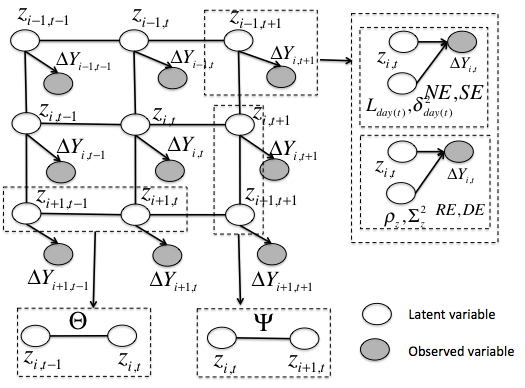}
\caption{Graphical Illustration of flu-MN for flu detection}\label{8png}
\end{figure}
\paragraph{Epidemic Dynamics}
Next, we describe the characteristics for the dynamics of different phases.
For {\bf NE} and {\bf SE}, the dynamics would only be affected by {\it daily effect}, where $\Delta Y_{i,t}$ is characterized as Gaussian process:
\begin{equation}
Pr(\Delta Y_{i,t}|z) \sim N(L_{day(t)}, \delta_{day(t)}^2)
\end{equation}

For {\bf RE} and {\bf DE} phases, the changes for epidemic dynamic would be great and {\it inter-related}. We model the conditional distribution of $\Delta Y_{i,t}$ as an autoregressive process combined with {\it daily effect}.
\begin{equation}
Pr(\Delta Y_{i,t}|z)\sim N(\rho_{z}+L_{day(t)}, \Sigma^2_{z}+\delta^2_{day(t)})
\end{equation}
$\Sigma^2_{Z}$ denotes the variance for $\Delta Y_{i,t}$ and we give a {\it inverse $\chi^2$} prior for them. $\rho$ is the autoregressive parameter {\it rising} and {\it declining} process, and we give a uniform prior for them.
\begin{equation}
\Sigma^2_{z}\sim Inv-\chi^2(\epsilon_{z},\psi^2_{z})
\end{equation}
\begin{equation}
\rho_{z=1}\sim Uniform(0,2)~~~~\rho_{z=3}\sim Uniform(-2,0)
\end{equation}

\subsection{Inference}
Along standard lines for Bayesian graphical models, the joint density of the variables are described as follows:
$[\cdot]$ denotes a probability density and $[\cdot|-]$ denotes the conditional density of parameter $(\cdot)$ given all the
other unknowns. We adopt Gibbs Sampling based on Markov Chain Monte Carlo where each latent parameter is sampled from posterior distribution given other parameters. 
\par
\noindent {\bf Sampling $Z_{i,t}$}:
\begin{equation}
\small
\begin{aligned}
&[Z_{i,t}|-]\propto [Z_{i,t}|Z_{i,t-1}, Z_{i,t+1},Z_{j,t}, j\in Neigh(i)]\cdot [\Delta Y_{i,t}|Z_{i,t}]\\
&\propto \frac{exp(\Theta_{z_{i,t-1},z_{i,t}}+\Theta_{z_{i,t},z_{i,t+1}}+\sum_{j\in Neigh(i)}\Psi_{z_{j,t},z_{i,t}})}{\sum_{z_{i,t}} exp(\Theta_{z_{i,t-1},z_{i,t}}+\Theta_{z_{i,t},z_{i,t+1}}+\sum_{j\in Neigh(i)}\Psi_{z_{j,t},z_{i,t}})}\\
&\times  [\Delta Y_{i,t}|Z_{i,t}]\\
\end{aligned}
\end{equation}

\noindent {\bf Sampling $L_k$ and $\delta^2_k$}:
$L_k$ and $\delta_k^2$ are updated at {\bf NE} and {\bf SE} phases:
\begin{equation}
\small
[L_k|-]=[L_k]\cdot \prod_{\substack{Z_{i,t}=0,2\\day(t)=k}}[\Delta Y_{i,t}|\delta^2_k,L_k]\propto N(\frac{\frac{M_k}{\delta_k^2}+
\frac{W_k}{\phi_k^2}}{\frac{N_k}{\delta_k^2}+\frac{1}{\phi_k^2}},\frac{1}{\frac{N_k}{\delta_k^2}+\frac{1}{\phi^2_k}})
\end{equation}

\begin{equation}
\small
\begin{aligned}
&[\delta^2_k|-]=[\delta^2_k]\prod_{\substack{Z_{i,t}=0,2\\day(t)=k}}[\Delta Y_{i,t}|\delta^2_k,L_k] \propto Inv-\chi^2(\mu_k+N_k,\frac{\mu_0\delta_k^2+V_k}{\mu_k+N_k})
\end{aligned}
\end{equation}
where $N_k=\sum_{i,t}{\bf I}(Z_{i,t}=0,2; day(t)=k)$, $W_k=\sum_{i,t}{\bf I}(Z_{i,t}=0,2; day(t)=k)\Delta Y_{i,t}$ and $V_k=\sum_{i,t}{\bf I}(Z_{i,t}=0,2; day(t)=k)(\Delta Y_{i,t}-L_k)^2$. {\bf I()} is the indicator function.

\noindent {\bf Sampling $\Theta$ and $\Psi$}:\par
\begin{equation}
\small
\begin{aligned}
&[\Theta_{z,z'}|-]\sim [\Theta_{z,z'}]\cdot [Z|\Theta,\Psi]\propto \exp(-\frac{\Theta_{z,z'}}{2\sigma_{z,z'}^2})\\
&\times P(Z_{i,j}=z'|Z_{i,t-1}=z,\Theta,\Psi)^{\sum {\bf I}(Z_{i,t-1}=z',Z_{i,t}=z)}\\
&\times P(Z_{i,j}\neq z'|Z_{i,t-1}=z,\Theta,\Psi)^{\sum {\bf I}(Z_{i,t-1}=z',Z_{i,t}\neq z)}\\
\end{aligned}
\end{equation}
Since the posterior distribution for $\Theta$ does not look like distribution we know (no conjugacy), we apply a {\it Metropolis-Hashings} algorithm\footnote{\url{http://en.wikipedia.org/wiki/Metropolis_Hastings_algorithm}} for $\Theta$ sampling. Similar approach is used for $\Psi$.

\noindent {\bf Sampling $\Sigma^2$}:\par
\begin{equation}
\small
\begin{aligned}
&[\Sigma^2_{Z}|-]\propto [\Sigma^2_{Z}]\times [\Delta Y_{i,t}|Z_{i,t}=Z,\Sigma^2_Z,\epsilon_Z]=\\
&\propto Inv-\chi^2(\epsilon_{z}+N_z, \frac{\epsilon^2_{z}\psi_z^2+\sum_{i,t}{\bf I}(Z_{i,t}=z)(\Delta Y_{i,t}-\rho_z)^2}{\epsilon_{z}+N_z})
\end{aligned}
\end{equation}
where $N_{Z}=\sum_i\sum_t{\bf I}(Z_{i,t}=Z)$

\noindent {\bf Sampling $\rho_z$}:\par
\begin{equation}
[\rho_z|-]\propto [\rho]\prod_{Z_{i,t-1}=z\atop Z_{i,t}=z}[\Delta Y^w_{i,t}|\rho_{z},\sigma^2_z,\epsilon_Z, L_k]
\end{equation}
Due to the space limit, we skip the detailed mathematical derivation for posterior distribution. We run 1000 iterations for MCMC algorithm.

\section{Dataset Creation}
\label{experiments-1}
\subsection{Collecting Data}
\begin{figure}
\includegraphics[width=3in]{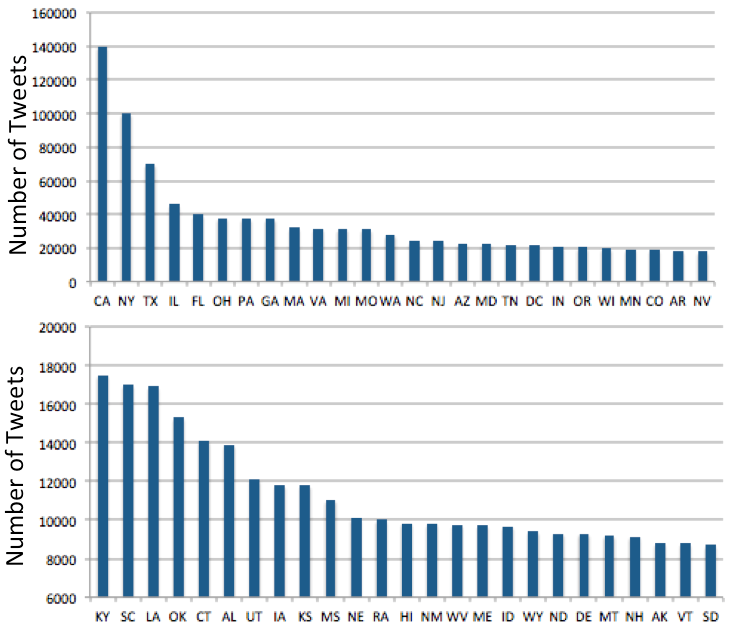}
\caption{State distribution of tweets in our datasets.}\label{fig1000}
\end{figure}

Based on the Twitter API, we developed a crawler to fetch data at regular time intervals. We fetched tweets containing indicator words shown in Table \ref{table1} and collect 3.6 million flu-related tweets from 0.9 million twitter users starting from Jun. 2008 to Jun. 2010. 
Location details can be obtained from the profile page or mobile client. We select tweets whose locations are in USA and discard those ones with meaningless locations such as "Mars", "in my bed" or "in the universe".  Olausible locations of users are then inferred by using Google Geocoding\footnote{\url{https://developers.google.com/maps/documentation/geocoding/}} by first converting the address into geographic coordinates of latitude and
longitude and then rematching the coordinates to the name of States\footnote{3G enabled mobile clients to make his or her location public if he wishes to, which is also helpful in this work.}. From the data we crawled, $19\%$ of the total number of tweets are with exact locations. The State-wise distribute of tweets are shown in Figure \ref{fig1000}.

\begin{table}[!ht]
\centering
\begin{tabular}{|c|c|}\hline
Indicator words&flu, influenza, H5N1, H5N9, swine flu, bird flu\\\hline
\end{tabular}
\caption{Indicator words for data collection.}\label{table1}
\end{table}

\subsection{Identifying flu-indicator tweets}
Not all tweets containing indicator keywords indicate that the user is ill. Twitter dataset needs to be processed before be used for our task.
The necessity of tweets filtering in real-time task has been demonstrated in many existing works \cite{aramaki2011twitter,lamb2013separating,sakaki2010earthquake}. 
.

Tweets can include 
mentions of target events. A user might post tweets such as "not feeling well, maybe got a flu" when he indeed has got a flu.
But not all tweets that include mentions of {\it influenza} or {\it flu} are true {\it flu indicators} 
indicating that the Twitter user has been affected.
For example, 
a user may say "Starting to get worried about swine flu" to express his concerned awareness, or "Swine flu vaccine is not good for pupils" to report the news. In addition,
even if a tweet is referring to a flu breakout, it might not be appropriate to treat it as an event report. For instance, a user may post tweets
such as "the bird flu last year is scaring", describing an out-of-date event, but not a real-time one. These kinds of inaccurate or biased information would have negative effect in our task.

To filter out these bias tweets, we first prepared manually labeled training data, which was comprised of 5000 tweets containing key words such as ``flu" or ``influenza".
We manually annotate them as positive examples 
(flu indicators) and negative ones (not indicative of the flu). We intentionally included the following cases in negative examples:
\begin{itemize}
\setlength{\itemsep}{0pt}
\item Starting to get worried about swine flu. [{\bf Awareness}].
\item Swine flu vaccine is not only for pupils. [{\bf Related News}]. 
\item The bird flu last year is scary. [{\bf Out of date}].
\item My son got flu, really worried. [{\bf Referring Another person}].
\item Do you think swine flu will come to America? [{\bf Question}].
\item Doctors urged people at risk of contracting swine flu to get vaccinated. [{\bf Warning}].
\end{itemize}
We built a classifier based on support vector machine.
We use SVMlight \cite{joachims1999making} with a polynomial kernel.
We employ the following simple text-based features:
\begin{description}
\setlength{\itemsep}{0pt}
   \item [Feature A]: Collocational features, representing words before and after the query word within a window size of three.
   \item [Feature B]: unigrams, denoting the presence or absence of the terms from the dataset.
   \item [Feature C]: tweet length in tokens.
   \item [Feature D]: position of the keyword within the tweet 
\end{description}

Performance for different combinations of features (shown at Section 2.1) are illustrated at Table \ref{table:svm}. We observe that the performance of A+B and A+B+D are much better than A+B+C and A+B+C+D. This illustrated that the consideration of tweet length (feature C) would largely affect the performance of classifier. We can also find that the position of keyword (feature D) does not contribute much to the classification performance. So in our following experiments, tweets are selected according to a classifier based on feature A+B.
\begin{table}[!ht]
\centering
\begin{tabular}{cccc}
Features&Accuracy&Precision&Recall\\\hline
A&84.41$\%$&83.31$\%$&91.40$\%$\\
B&85.40$\%$&86.94$\%$&88.00$\%$\\
A+B&87.30$\%$&88.65$\%$&89.54$\%$\\
A+B+D&87.26$\%$&88.35$\%$&89.85$\%$\\
A+B+C&74.49$\%$&73.61$\%$&87.21$\%$\\
All&73.04$\%$&71.75$\%$&88.16$\%$\\\hline
\end{tabular}
\caption{Performance for different combinations of features for tweets filtering}\label{table:svm}
\end{table}

\subsection{Further Tweets Selection}
For later analysis, tweets dataset needs to be further processed to filter redundant tweets \cite{ab,achrekar2011predicting}. Issues are explained as follows:
\begin{itemize}
\item {\bf Retweets}: A retweet is a tweeted post originally made by one user and then followed or forward by someone else. For flu tracking, it does not indicate a new user being affected. Retweets would be filtered out in our system.
\item {\bf Duplicate Tweets}: An individual may publish more than one tweets talking about his symptoms when he got a flu. 
He may first tweet "seems like caught a flu", then "feel bad, got a flu" and at last "finally recovered from the flu".
To get rid of duplicating tweets, we remove tweets from the same user within the same syndrome time (two weeks). 
\end{itemize}


\subsection{Comparing Tweet Trend with CDC}
\begin{figure}[!ht]
\centering
\hspace{-0.5cm}
\includegraphics[ width=3in,natwidth=610,natheight=642]{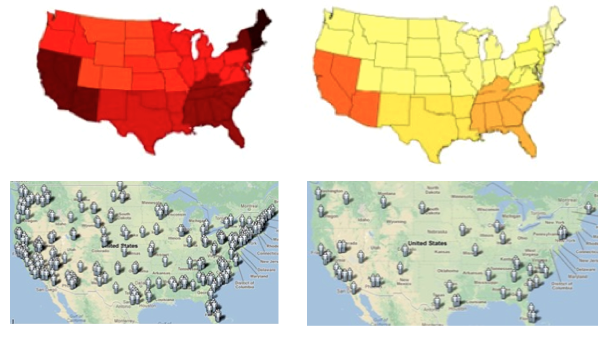}
\caption{Visual illustration of CDC data and flu-related tweets distribution.The upper CDC data pictures are from \cite{ab}}
\label{visual}
\end{figure}

In this subsection, we briefly demonstrate the relatedness between Twitter data and CDC data, which would support the claim that Twitter data can be used for the flu detection task. We quantitatively evaluate the correlated performance between CDC data and twitter data by reporting $correlation$ and {\it RMSE error}\footnote{\url{http://en.wikipedia.org/wiki/Root_mean_square_deviation}} in Table (\ref{table2})\footnote{Due to the space limit, we only report the results according to regions.}. We observe that performing SVM filtering and tweets selection would definitely make twitter data more correlated with real world CDC data. 
Figure \ref{visual} presents the distribution map published by CDC in week 45 (Nov15-Nov21), 2009 (left column) and week 6 (Feb7-Feb13), 2010 (right column) and the correspondent visualization of tweets containing keywords according to geo-location . In left column of Figure \ref{visual}, we can clearly identify the states with big clusters of flu related tweets including California,  Arizona, Florida, Georgia and most states in New England, which are highly correlated with CDC reports. These results illustrate the high correlation between Twitter data and clinical data(CDC), which enables twitter data to be used for early-stage flu detection in the real world.  

\begin{table}
\scriptsize
\centering
\begin{tabular}{|c|c|c|c|c|}\hline
\multirow{2}{*}{}&\multicolumn{2}{c|}{non-processing}&\multicolumn{2}{c|}{processing}\\\hline
& Correlation& RMSE errors& Correlation&RMSE errors\\\hline
United States&0.965&0.372&{\bf 0.974}&{\bf 0.332}\\
New England&0.963&0.369&{\bf 0.972}&{\bf 0.319}\\
Mid-Atlantic&0.970&0.375&{\bf 0.978}&{\bf 0.308}\\
East-North Central&{\bf 0.958}&0.392&{\bf 0.958}&{\bf 0.364}\\
West-North Central&0.982&0.343&{\bf 0.984}&{\bf 0.317}\\
Mountain&0.973&0.402&{\bf 0.981}&{\bf 0.388}\\
Pacific&0.964&0.377&{\bf 0.982}&{\bf 0.358}\\
West South Central&0.962&0.368&{\bf 0.977}&{\bf 0.320}\\
East South Central&0.974&0.334&{\bf 0.982}&0.312\\
South Atlantic&0.966&0.372&{\bf 0.980}&{\bf 0.340}\\\hline
\end{tabular}
\caption{Correlation between twitter data and CDC data. non-processing corresponds to approach without SVM filtering and further tweets selection.}\label{table2}
\end{table}

\section{Experiments}
\label{experiments-2}
In this subsection we present experimental results  for early stage flu detection based on the formulation described in Section 2. 
\subsection{Parameter Evaluation}
We first discuss parameters estimated from Flu-MN. 
Figures \ref{12}(a) and (b) respectively show the visualization of estimations for parameter $\Theta$ and $\Psi$. $\Theta$ can be interpreted as how likely phase $i$ takes place at time $t$ given that phase $j$ appeared at the previous time $t-1$. We observe that a {\bf NE} phase has very low probability to transit to {\bf DE},  and likewise for the transiting probability from {\bf SE} to {\bf NE} and {\bf DE} to {\bf RE}, which agree with general rules. 
To better illustrate this point, in Table \ref{table100}, we further present the value of  $Pr(z_{i,t}|z_{i,t-1})$ and $Pr(z_{i,t}|z_{j,t})$ by integrating out other parameters in Equ.4. As we can observe, the model estimates that each location is very likely to stay at the phase it took on during the previous time and neighboring locations tend to share similar phases.

\begin{figure}[!ht]
\centering
\includegraphics[width=3.4in]{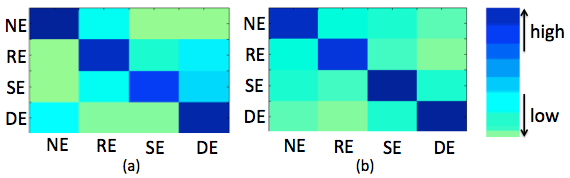}
\caption{Visualization of (a) $\Theta$: Temporal Dependency (b) $\Psi$: Spatial Dependency}\label{12}
\end{figure}

\begin{table}[!ht]
\centering
\includegraphics[width=3.6in]{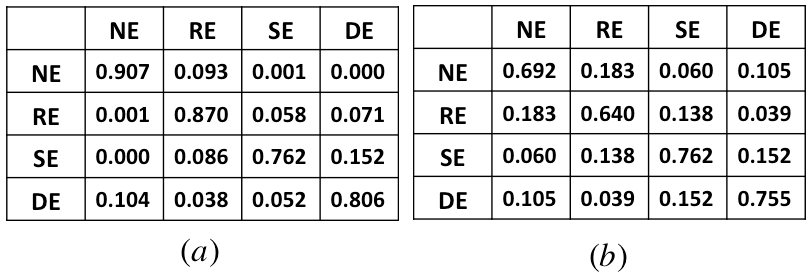}
\caption{Value for (a)$Pr(z_{i,t}|z_{i,t-1})$ and (b)$Pr(z_{i,t}|z_{j,t})$}
\label{table100}
\end{table}

Table \ref{table3} shows the posterior mean value of parameters involved in our model and correspondent $95\%$ confidence interval.
We observe that for weekdays (i.e. $L_2,L_3,L_4,L_5$), {\it daily effect} tend to be characterized by small random changes around zero, in other words, a {\it Gaussian white noise process}. 
Users tend to publish more flu-related tweets on weekdays than weekends, which can be explained by the fact that users tend to have larger to be in close contact with others in public places or work
, hence easier to get affected. There would be major fluctuation for the number of flu-related tweets published on Saturday, Sunday and Monday, leading to larger value (absolute value) of $L_1,L_2$, and $L_7$.
We also see that $\delta_k^2$ are smaller than the variances of epidemic phase $\Sigma^2_{z=1}$ and $\Sigma^2_{z=3}$ , denoting the minor fluctuation for static phase, in accordance with common sense.

\begin{table}[!ht]
\scriptsize 
\centering
\begin{tabular}{|ccc|ccc|}\hline
{\small}&{\small posterior}&$\multirow{2}{*}{} 95\%$ confidence&{\small}&{\small posterior}&$\multirow{2}{*}{}95\%$ confidence\\
&&interval&&&interval\\\hline
$\rho_{z=1}$&0.721&[0.482,0.960]&$\rho_{z=1}$&-0.572&[-0.782,-0.360]\\
$\Sigma^2_{z=1}$&0.467&[0.344,0.590]&$\Sigma^2_{z=3}$&0.398&[0.342,0.454]\\
$L_1$&-0.210&[-0.524,0.104]&$\delta_1^2$&0.087&[0.073,0.101]\\
$L_2$&0.607&[0.420,0.794]&$\delta_2^2$&0.142&[0.117,0.167]\\
$L_3$&0.110&[-0.002,0.222]&$\delta_3^2$&0.011&[0.007,0.015] \\
$L_4$&-0.092&[-0.205,0.021]&$\delta_4^2$&0.013&[0.009, 0.018]\\
$L_5$&0.011&[-0.102,0.124]&$\delta_5^2$&0.015&[0.011,0.019]\\
$L_6$&-0.004&[-0.082,0.074]&$\delta_6^2$&0.010&[0.008,0.012]\\
$L_7$&-0.312&[-0.102,0.520]&$\delta_7^2$&0.052&[0.041,0.063]\\\hline
\end{tabular}
\caption{Posterior of the parameters in our model.}\label{table3}
\end{table}

\begin{figure*}[!ht]
\centering
\subfigure{
\includegraphics[ width=6in,natwidth=610,natheight=642]{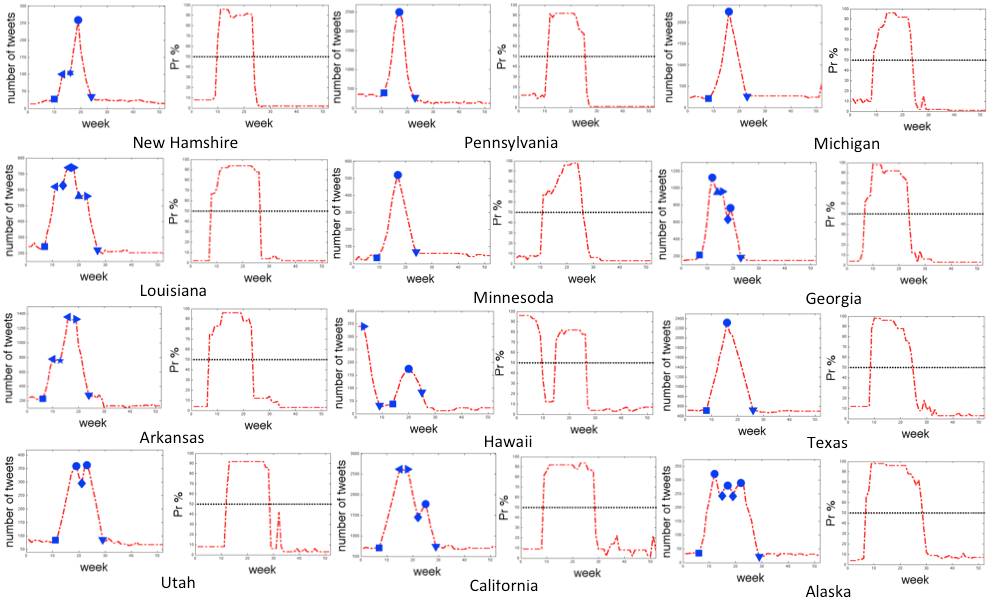}}
\subfigure{
\includegraphics[ width=6in,natwidth=610,natheight=642]{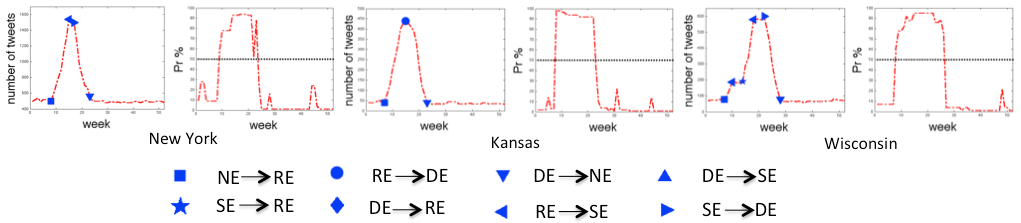}}
\caption{Examples of our model on the whole year Twitter data from different States}.\label{fig10}
\end{figure*}

\hspace{-1.5cm}
\begin{table*}[!ht]
\scriptsize 
\centering
\begin{tabular}{|cccccc|cccccc|}\hline
State&Peak Time&flu-MN&Average&HMM&Two Phase&State&Peak Time&flu-MN&Average&HMM&Two Phase\\\hline
Oregon&Oct.01&{\bf Aug.14}&Sep.02&Aug.25&Aug.22&Washington&Oct.04&{\bf Aug.08}&Aug.28&Aug.12&Aug.22\\
Idaho&Sep.23&{\bf Aug.08}&Aug.25&Aug.17&Aug.12&Arkansas&Sep.08&{\bf Jul.14}&Aug.14&Jul.18&Jul.16\\
Utah&Oct.06&{\bf Jul.26}&Aug.18&Aug.01&Aug.04&California&Sep.18&{\bf Aug.01}&Aug.18&{\bf Aug.01}&Aug.06\\
Arizona&Sep 26&{\bf Aug.08}&Aug.28&Aug12&Aug.14&New Mexico&Oct.01&{\bf Aug.10}&Sep.12&Aug.22&Aug.18\\
Montana&Oct.12&{\bf Aug.18}&Sep.16&Aug.24&Aug.28&Wyoming&Oct.06&{\bf Aug.04}&Sep.10&Aug.12&Aug.14\\
Colorado&Sep.30&{\bf Aug 03}&Aug.31&Aug.14&Aug.12&Minnesota&Oct.22&{\bf Aug.21}&Oct.01&Aug.27&Aug.22\\
Kansas&Oct.01&{\bf Jul.30}&Aug.22&Aug.08&{\bf Jul.30}&Oklahoma&Sep.20&{\bf Aug.06}&Aug.16&Aug.16&Aug.09\\
Texas&Sep.24&{\bf Aug.14}&Sep.4&Aug.16&Aug.18&Tennessee&Aug.24&{\bf Jul.12}&Jul.26&Jul.18&Jul.17\\
Florida&Sep.12&{\bf Jul.28}&Aug.16&{\bf Jul.28}&Aug.02 &Mississippi&Aug.28&{\bf Jul.09}&Aug.01&Jul.22&Jul.24\\
Iowa&Oct.09&{\bf Aug.30}&Sep.12&Sep.11&Sep.08&Wisconsin&Oct.19&{\bf Sep.06}&Sep.18&Sep.07&Sep.08\\
Nevada&Sep.14&{\bf Aug.15}&Sep.04&Aug.25&Aug.22&Kentucky&Aug.30&{\bf Jul.19}&Aug.01&Jul.28&Jul.28\\
Michigan&Oct.08&{\bf Aug.14}&Aug.29&Aug.27&Aug.22&New York&Oct.18&{\bf Aug.26}&Sep.28&Sep.01&Sep.08\\
Ohio&Oct.04&{\bf Aug.06}&Aug.20&{\bf Aug.06}&Aug.08&Pennsylvania&Oct.16&{\bf Aug.24}&Sep.16&Sep.04&Sep.06\\
Indiana&Oct.07&{\bf Aug.12}&Oct.01&Aug.17&Aug.17&West Virginia&Oct.12&{\bf Aug.17}&Sep.02&Aug.24&Aug.28\\
Georgia&Sep.17&{\bf Jul.28}&Aug.15&Aug.02&Aug.06&South Dakota&Oct.05&{\bf Aug.15}&Sep.16&Aug.21&Aug.24\\
Maine&Oct.16&{\bf Aug.22}&Sep.14&Aug.28&Aug.30&New Hampshire&Oct.16&{\bf Aug.22}&Sep.18&Aug.28&Sep.03\\
Vermont&Oct.18&{\bf Aug.21}&Sep.12&Aug.26&Aug.30&Massachusetts&Oct.12&{\bf Aug.20}&Aug.29&Aug.28&Aug.26\\
Delaware&Oct.14&{\bf Aug.16}&Sep.18&Aug.29&Aug.30&New Jersey&Oct.16&{\bf Aug.24}&Sep.12&Sep.05&Sep.01\\
Maryland&Oct.08&{\bf Aug.12}&Sep.06&Aug.18&Aug.22&Connecticut&Oct.18&{\bf Aug.16}&Sep.09&Oct.22&Oct.25\\
Alabama&Sep.18&{\bf Jul.25}&Aug.14&Aug.04&Aug.07&North Carolina&Oct.01&{\bf Aug.02}&Aug.26&Aug.08&Aug.04\\
Louisiana&Sep.20&{\bf Jul.28}&Aug.10&Sep.29&Sep.27&South Carolina&Oct.06&{\bf Aug.16}&Sep.01&Aug.22&Aug.26\\
Virginia&Oct.14&{\bf Aug.14}&Sep.12&Aug.18&Aug.24&Rhode Island&Oct.16&{\bf Aug.22}&Sep.09&Aug.29&Aug.28\\
Missouri&Sep.10&{\bf Jul.18}&Aug.26&Jul.28&Jul.31&Columbia&Oct.06&{\bf Aug.16}&Sep.06&Aug.26&Aug.27\\
Hawaii&Jun.16&{\bf Apr.18}&May.16&{\bf Apr.18}&Apr.25&North Dakota&Oct.03&{\bf Aug.11}&Sep.12&Aug.18&Aug.18\\
Alaska&Sep.20&{\bf Aug.16}&Sep.04&{\bf Aug.16}&Aug.26&Nebraska&Sep.24&{\bf Aug.05}&Aug.29&Aug.12&Aug.10\\\hline
\end{tabular}
\caption{Comparison with different for real-time phase prediction detection from Jun, 2009 to Jun. 2010.}\label{table:baselines}
\end{table*}

\subsection{Flu Detection from Real-time Data}
The main goal of our task is to help government or health organizations to raise an alarm at those moments
when there is a high probability that the flu breaks out. In real time situations, for each time, available data
only comes from the previous days, and there is no known information about what will happen in the following days or weeks.
By adding the data day by day, we calculate the posterior probability for transiting to (or staying at)
epidemic states (including {\bf RE}, {\bf SE} and {\bf DE}) based on previous observed data. 
The prediction for phase $Z_{i,t}$ would be attained by maximizing the likelihood given $Y_{i,t}$ as follows:
\begin{equation}
\begin{aligned}
&Z_{i,t}=\argmax_{z} P(Z_{i,t}=z|Y_{i,t},Y_{i,t-1},...)\\
&=\argmax_{z}\sum_{\substack{Z_{i,t-1}\\Z_{i,t-1},\\...}} P(Z_{i,t}=z|Z_{i,t-1},Z_{i,t-2},..., Y_{i,t},Y_{i,t-1},...)\\
\end{aligned}
\label{12}
\end{equation}
The sum over parameter $Z_{i,t-1}$ and $Z_{j,t}$ makes it infeasible to calculate.  We again use Gibbs Sampling by first sampling $Z_{i,t-1}$ and $Z_{j,t}$ first and then attain the value of $Z_{i,t}$ given $Z_{i,t-1}, Z_{i,t-2}, ...$:
\begin{equation}
Z_{i,t}=\argmax_{z} P(Z_{i,t}=z|Z_{j,t},Z_{i,t-1},..., Y_{j,t},Y_{i,t-1},...)\\
\end{equation}
Figure \ref{fig10} shows the performance of our real-time analysis based on twitter data from Jun.2009 to Jun.2010. 
We randomly select $15$ States in United States.
For each State in Figure \ref{fig10}, the left hand side figure corresponds to number of flu-related tweets over time. Blue symbols denote the phase transition point detected by our approach from non-epidemic phase (NE) to epidemic phase (RE,NE or DE).
The right hand side figure corresponds to posterior probability of being at epidemic phase. The dotted line denotes Pr(epidemic)=0.5 when the alarm would be sent out.
Note that these probabilities can be very useful for government and organizations. 
In particular, value exceeding 0.5 indicates that at that time, the State witnesses a higher probability being at epidemic phase than non-epidemic one. The black dotted line represents $Pr(epidemic)=0.5$, where there should be a flu alarm \cite{martinez2008bayesian}. 
Another interesting thing we observe from Figure \ref{fig10} is that jumps from one phase to another are quite robust in our model. There is no more than one epidemic transition from {\bf NE} to {\bf RE} (i.e. non-epidemic phase to epidemic phase) 
per season even though the model allows for any number of changes. This further demonstrates the robustness of our model. 

 For comparison, we employ the following baselines in this paper: 

 {\bf Average}: An approach that uses the average frequency of tweets containing keywords based on previous yearn as the threshold. Days with larger frequency of tweets would be treated as epidemic. 
 
 {\bf Flu-MN(r)}: A revised version of Flu-MN where {\it daily effect} is not considered. The dynamics for {\bf NE} and {\bf SE} are modeled as a {\it white Gaussian noise process}, the expectation of which is 0.
 
{\bf Time HMM}: A simplified version of flu-MN where temporal dependency is modeled as $four-phase$ Markov chain but spatial dependency is neglected. 

{\bf Two-Phase}: A simple version of our approach but using a simpler two-phase (epidemic phase and non-epidemic phase) in Markovian network. Both spatial and temporal dependency is considered.

In Table \ref{table:baselines}, we present the performance from different baselines along with Flu-MN based on 50 States twitter data in Unites States from Jun 2009 to Jun 2010\footnote{Since most of phase transiting times detected by Flu-MN would also detected by Flu-MN(r) (but Flu-MN(r) involve more fluctuation and wrongly detected phase transiting times due to the neglect of {\it daily effect}), we do not report results for Flu-MN(r) here.}.
Peak time denotes the moment when the tweet frequency containing keywords reaches the peak. 
From Table \ref{table:baselines}, we can clearly identify that for most states, {\it Flu-MN} would detect the breakout of influenza earlier than {\it Time HMM}. This validates our assumption that by considering neighboring influence, we can achieve a better flu prediction performance. 
It is also worth noting that, for some States (i.e. Hawaii, Alaska, Florida), which have very few (or no) bordering States, {\it Flu-MN} would degenerate into {\it Time HMM}. That is why for these States, {\it Flu-MN} and {\it HMM} achieve very similar results. {\it Flu-MN} detects the phase transition time more promptly than {\bf two-state}, illustrating the effectiveness of using a more 
sophisticated four-phase framework to model flu dynamics than a simpler {\it two-state} one. Table \ref{109} illustrates the average  days that each model predicts the flu breakout before the number of flu-related tweets reaches its peak. 
\begin{table}
\small
\centering
\begin{tabular}{|c|c|c|c|c|c|}\hline
&Peak&Flu-MN&Average&Time HMM&Two-Phase\\\hline
Days&0&{\bf -48.6}&-25.8&-36.0&-34.4\\\hline
\end{tabular}
\caption{DIC value in regard to different baselines.}\label{109}
\end{table}

Further, in order to quantitatively compare competing models, we report deviation information criterion (DIC) at Table \ref{109}. DIC is  defined as a classical estimate of fit, plus twice the effective number of parameters $p_D$, the details of which can be found in Spiegelhalter et al.'s work \cite{spiegelhalter2002bayesian}. Flu-MN outperforms (by providing a better fit, which can be seen by a lower DIC) the other three models. 
\begin{table}
\centering
\begin{tabular}{|c|c|c|c|c|}\hline
&Flu-MN&Flu-MN(r)&Time HMM&Two-Phase\\\hline
DIC&{\bf 2441}&2588&2742&2767\\\hline
\end{tabular}
\caption{DIC value in regard to different baselines.}\label{109}
\end{table}

\subsection{Future Prediction}
\begin{table}[!ht]
\footnotesize
\centering
\begin{tabular}{|c|c|c|c|c|}\hline
&Flu-MN&Time HMM&Two-Phase&Auto-regress\\\hline
Correlation&{\bf 0.974}&0.949&0.947&0.917\\\hline
RMSD error&{\bf 0.317}&0.335&0.329&0.372\\\hline
\end{tabular}
\caption{Evaluation for future $Y$ prediction.}\label{table11}
\end{table}
Given observants from previous days, people are interested in asking questions like how likely flu will breakout tomorrow, or how likely flu will come to an end next week, or how many flu-related tweets there would be next week.  
We can use a similar approach we take at Section 5.2 for $Y_{i,t+1}$ and $Z_{i,t+1}$ prediction. Suppose that we have already attained the values of $Z_{i,t}, Z_{i,t-1}, ...$ from a Gibbs Sampling. Then, $Y_{i,t+1}$ and $Z_{i,t+1}$ can be predicted as follows:
\begin{equation}
Z_{i,t+1}=\argmax_{Z}P(Z_{i,t+1}=z|Y_{i-1,t}, ...)
\end{equation}
\begin{equation}
Y_{i,t+1}=\argmax_{Y}\sum_{Z_{i,t+1}=z}P(Y|z) P(Z_{i,t+1}=z|Z_{i-1,t}, ...)
\end{equation}

For comparison, in addition to {\it Time HMM} and {\it Two-Phase} for future $Y_{i,t+1}$ prediction, we employ the following baseline:

 {\bf Auto-regression}: An approach for $Y_{i,t}$ prediction based on $Y_{i,t-1}$, $Y_{i,t-2}$, ... based on regressive model. 
 \begin{equation}
 Y_{i,t}=\sum_{j=1}^{n}b_jY_{i,t-j}+c+e(i,t)
 \end{equation}
where $e(i,t)$ is a sequence of independent random variables and c is a constant term to account for the offset. Parameters are trained on tweet data from Jul.2008 to Jul.2009.

\begin{figure}
\centering
\subfigure{
\includegraphics[width=2.9in]{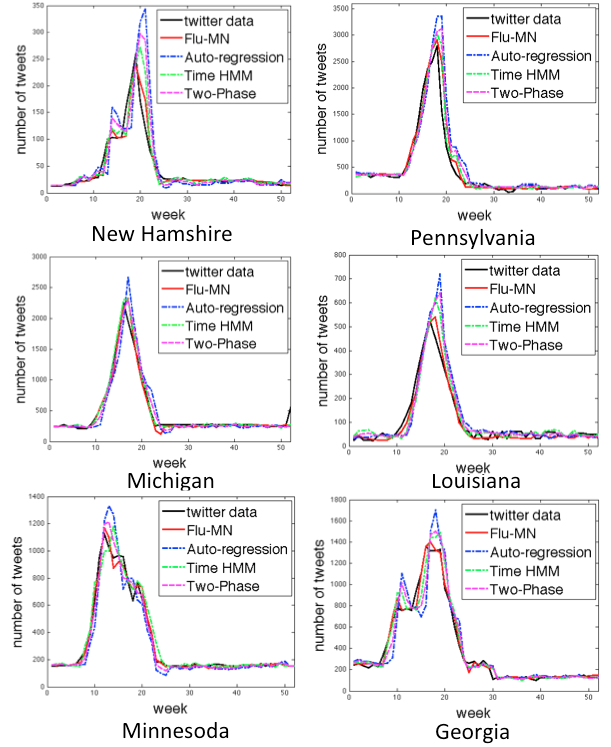}}

\subfigure{
\includegraphics[width=2.9in]{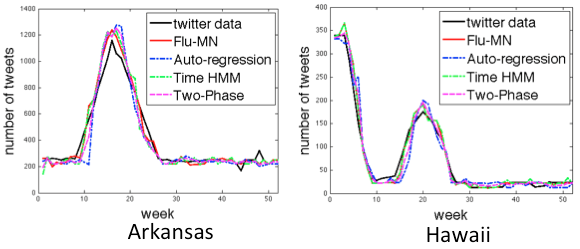}}
\caption{$Y$ predictions from Jun.2009 to Jun.2010. }\label{16}
\end{figure}

Figure \ref{16} presents $Y_{i,t+1}$ prediction performance compared with CDC data from Jul.2009 to Jul.2010. Due to space limit, we only report performances for 8 States. As we can see, {\it Flu-MN} (the red curve) can best fit the actual twitter data (the black 
curve). {\it Time HMM} and {\it Two Phase} attain similar performances. {\it Auto-regression} achieves the worst results. This can be explained as follows: flu trends vary quite a lot from each other in different years. So parameters trained from data in 2008-2009 may not be well tailored for 2009-2010. For a better illustration, quantitative evaluations are reported at Table \ref{table11}.

\begin{figure}[!ht]
\centering
\subfigure{
\includegraphics[width=2.8in]{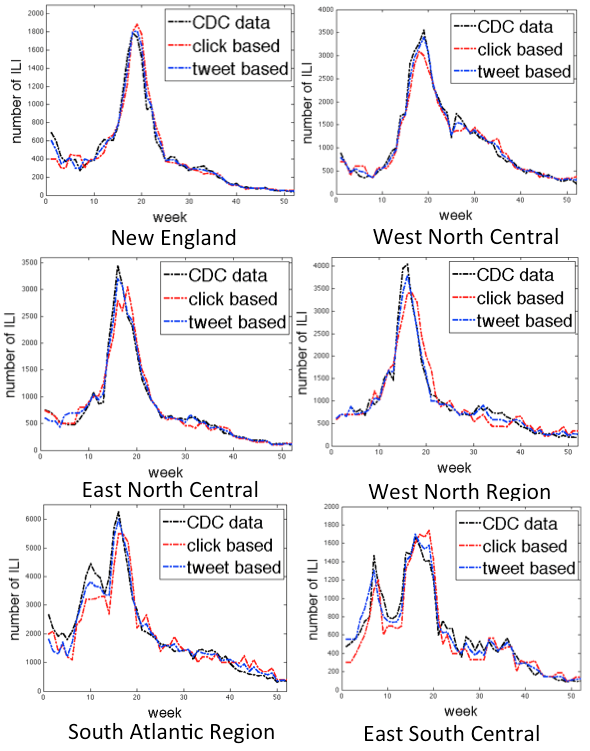}}
\subfigure{
\includegraphics[width=2.8in]{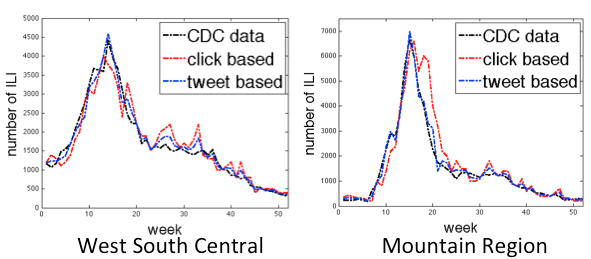}}
\caption{ILI prediction from Jun.2009 to Jun.2010. }\label{190}
\end{figure}

\subsection{Twitter or Google Flu Trend?}
Another interesting task involved is the prediction of the ILI related physician visits, which, actually, is a main task addressed in Google Flu trend \cite{ginsberg2008detecting}.
Here, we adopt the linear model using log-odds which is used in Ginsberg et al.'s work \cite{ginsberg2008detecting} to link the number of ILI physician visits with the number of flu-related tweets. Let $N^{ILI}_{i,t}$ be the number of ILI visits\footnote{CDC only provides weekly data according to regions, we focus on weekly regional ILI prediction.}. We have:
\begin{equation}
\log(N^{ILI}_{i,t})=\beta_0+\beta_1\cdot \log(Y_{i,t})+\epsilon
\end{equation}
where $\beta_0$ is the intercept, $\beta_1$ is the multiplicative coefficient and $\epsilon$ is the error term. We fit the linear model using CDC data and flu-related tweet data from 2008-2009 and test the model based on actual CDC data from 2009-2010.

Figure \ref{190} presents a comparison between estimates of ILI physician visits by using tweet data and Google flu trend click which is based on online query search in different regions of the United States. As we observe, prediction based on flu-related tweets fits the actual CDC data netter than online query-search based approach. Table \ref{1900} also demonstrates this point. The better performance of twitter over click based methods is expected. By applying pre-processing techniques in Section 3, we can keep the tweets that exactly indicate user being affected and discard non-flu indicating ones. However, for online clicks, it's much harder to tell which type of click exactly indicates a user being affected. An direct example is that a conference on flu or epidemic may largely increase the frequency of flu related query searches, but it does not indicates a flu breakout. Moreover, flu related query clicks include multiple types (e.g. General Influenza Symptoms, Antibiotic Medication, Influenza Remedies), which are not necessarily related to a flu breakout (say Antibiotic Medication).
\begin{table}
\centering
\begin{tabular}{|c|c|c|}\hline
&Correlation&RMSE errors\\\hline
Twitter based&{\bf 0.958}&{\bf 0.364}\\
Query-search based&0.917&0.390\\\hline
\end{tabular}
\caption{Comparison between ILI prediction}\label{1900}
\end{table}

\section{Related Work}
\label{rel-work-sec}
\paragraph{Real-world event detection on Twitter}
Twitter serves as a good source for quick event detection due to its real time nature. As described in the introduction, Sakaki et al. \cite{sakaki2010earthquake} were the first to build an earthquake and typhoon detection system from Twitter data,  making the assumption that each Twitter user could be regarded as a sensor. 
They produced a probabilistic model that can find the center of the earthquake
and its trajectory and is able to send alerts
faster than meteorological agencies. 

\paragraph{Online Flu Surveillance}
Early work on web-based influenza
surveillance usually makes use of click-through
data from search engines \cite{eysenbach2006infodemiology,cook2011assessing,ginsberg2008detecting,hulth2009web,johnson2004analysis,polgreen2008using},
 such as using medical website with influenza \cite{hulth2009web}, 
counting visitors to health websites \cite{johnson2004analysis}. 
or users' clicks on a 
search keyword advertisement \cite{eysenbach2006infodemiology}. Google's flu trend is also a online-click based flu system that uses a linear model to link the number of ILI-related search query to the number of ILI-related physician visit.

Twitter has been proved to be useful for flu surveillance
\cite{achrekar2011predicting,achrekar2012twitter,ji2012epidemic,ritterman2009using,sadilek2012modeling}. Achrekar et al. \cite{achrekar2011predicting} 
studied the correlation between CDC data and flu-related tweets in twitter.
Aramaki et al. \cite{aramaki2011twitter} and Lamb et al. \cite{lamb2013separating} explored multiple features in supervised learning by classifying different kinds of tweets containing keywords, trying to select out tweets indicating flu around.

\paragraph{Flu Detection in Epidemiology}
Influenza detection and prediction problem can be traced back to Serfling's work \cite{serfling1963methods} in 1963 in epidemiology, which tried to find a threshold for influenza breakout. Since then, various approaches have been proposed for flu detection and prediction in multiple situations \cite{ab,christenson1982epidemic,groendyke2011bayesian,le1999monitoring,moreno2002epidemic,mugglin2002hierarchical}. LeStrat and Carrat \cite{le1999monitoring} segmented the time series of influenza indicators into epidemic and nonepidemic phases by incorporating Markovian Assumption and used an EM algorithm for parameter estimation.
Subsequently, Rath et al. \cite{rath2003automated} and Madigan \cite{madigan2005bayesian} presented further investigation of the model from the Bayesian perspective. As for current efforts, Muglin \cite{mugglin2002hierarchical} treat the disease fonts as a realization of multivariate autoregressive process and Sebastiani et al. \cite{sebastiani2006bayesian} integrate different sources of data into a multivariate model for influenza surveillance. 
 These approaches are mostly based on CDC or EISS datasets.

\section{Conclusion}
\label{conclusion-sec}
In this paper, we introduced a unsupervised Bayesian model based on Markov Network for early stage flu detection on Twitter. We test our model on real time datasets for multiple applications and experiments results demonstrate the effectiveness of our model. We are hopeful that our approach would help to facilitate timely action by government
and health organizations who want to decrease the number and cost of unnecessary illnesses and deaths.

\bibliographystyle{abbrv}
\bibliography{sigproc}  
\end{document}